# Three-dimensional flow with elevated helicity in driven cavity by parallel walls moving in perpendicular directions

Alex Povitsky

Professor, AIAA Associate Fellow, Email: povitsky@uakron.edu

Department of Mechanical Engineering, University of Akron, Akron, OH 44325-3903 USA

Abstract

The proposed flow in a 3-D cubic cavity is driven by its parallel walls moving in perpendicular directions to create a genuinely three-dimensional highly separated vortical flow yet having simple single-block cubical geometry of computational domain. The elevated level of helicity is caused by motion of a wall in the direction of axis of primary vortex created by a parallel wall. The velocity vector field is obtained numerically by using second-order upwind scheme and $200^3$ grid. Helicity, magnitude of normalized helicity and kinematic vorticity number are evaluated for Reynolds numbers ranging from 100 to 1000. Formation of two primary vortices with their axis oriented perpendicularly and patterns of secondary vortices are discussed. Computational results are compared to the well-known 3-D recirculating cavity flow case where the lid moves in the direction parallel to the cavity side walls. Also results are compared to the diagonally top-driven cavity and to cavity flow driven by moving top and side walls. The streamlines for the proposed flow show that the particles emerging from top and bottom of cavity do mix well. Quantitative evaluation of mixing of two fluids in the proposed cavity flow confirms that the mixing occurs faster than in the benchmark case.

1. Introduction

This study introduces a simple geometry model of genuinely three-dimensional and highly separated flow in cubical enclosed cavity with its top and bottom walls moving in orthogonal directions (Case A in Fig. 1) and studies the created flowfield as a function of the Reynolds number. Compared to prior set-ups of recirculating flow in driven cavity, the proposed flowfield has elevated helicity, sustained primary vortices with orthogonal axis of rotation, and two-sided mixing.

Technological applications include mixing devices including micro-fluidic mixers. A single screw polymer extruder for mixing of particles consists of an Archimedean screw rotating in a barrel [1]. The system can be modeled as a channel or cavity in which the upper plate (that is, barrel) is moved diagonally across the top with helix angle relative to the screw (see [1], Fig. 13). If, in addition to the moving screw, barrel rotates to enhance mixing, it creates flow similar to that in Case A. In particular, intensification of laminar flow mixing in micro-channels is important and various ways to enhance mixing were proposed including patterned grooves [2,3,4], that form semi-opened cavities [5]. To further enhance mixing, either the walls of channel or centrally located shaft could rotate. This would create motion of cavity bottom wall in the direction perpendicular to the channel axial flow motion that creates flow similar to Case A.

According to Ref. [6], internal recirculating cavity flows generated by the motion of one or more of the containing walls are not only technologically important, they are of great scientific interest because they display almost all fluid mechanical phenomena in the simplest of geometrical settings. Driven cavity flows offer an ideal framework in which meaningful and detailed comparisons can be made between results obtained from experiment, theory, and computation [6]. The current study introduces and investigates novel recirculating cavity flow with elevated helicity and mixing. By [7], the scalar product of velocity and vorticity, $\vec{u}\cdot\vec{\omega}$, is defined as the helicity per unit volume of the flow. The level of helicity in the proposed flow set-up appears to be larger than those for prior cavity flow set-ups (Cases B and C in Fig. 1) because of elevated helicity at near-wall layer and in vortices.

Helicity plays an important role in characterizing complex three-dimensional flows including mixing, loss of stability and transition to turbulence (see [8] and references therein), vortex breakdown, the growth of magnetic fields in electrically-conducting fluids (see review [9] and references therein) and topology of vortices [10]. Helicity can lead to better mixing of chemical components in helical coherent structures (see [11] and references therein). The Arnold–Beltrami–Childress (ABC) [12] flow is a three-dimensional incompressible velocity field which is an exact solution of inviscid Euler equations in which helicity is the maximum (velocity and vorticity vectors are parallel). The ABC flow is a simple inviscid example of a fluid flow that can have chaotic trajectories and intense mixing associated with substantial helicity. This calls for the need to create a simple geometric set-up of viscous flow with elevated helicity.

In addition, helicity is used as a key variable in CFD algorithms and therefore benchmark flows in simple geometry can be used for verification of numerical methods. For example, Ref [13] suggested the use of helicity as an independent variable for numerical solution of vortex-dominated three-dimensional flows. Normalized helicity is used for visualization of vortices shed by air vehicles [14].

In the current study, the topology of vortices and level of helicity are compared to the widespread 3-D lid-driven cavity flow (Case C in Fig. 1) for which data are available [6]. Case C is the 3-D extension of the well-known 2-D lid-driven cavity flow benchmark. The prior studies [15] introduced flows in 2-D cavities in which the top and bottom walls move either in the same direction or in opposite direction with same velocity. However, for 2-D cavity flows vorticity is perpendicular to the cavity plane, velocity and vorticity are orthogonal and therefore helicity is equal to zero. For Case C, the velocity and vorticity are nearly orthogonal at each (y, z) plane except for the cavity ends and therefore the level of helicity appears to be low compared to new Case A. Studies [16,17] show that the 3-D cavity flow (Case C) becomes unstable only when Re>2000. To limit the current study to stationary flows, the selected range of Reynolds numbers is below 2000.

The integral value of helicity for the proposed Case A is compared to that for the cubic cavity flow with diagonal symmetry in which the top wall moves along its diagonal (Case B-1 in Fig.1), Case B-1 was introduced by the author [18]. In fluids engineering, the Case B-1 flow set-up is important to model pneumatic transport of dusts [19] and to compute the degree of pollutant escape from the cavity according to various meteorological factors [20]. For verification and validation of numerical methods, the Case B flow set-up has been used to develop an artificial compressibility method for the incompressible Navier–Stokes equations [21], a preconditioned Krylov solver for stratified oceanic flows [22], multiple–relaxation–time Lattice-Boltzmann method, and to apply Lattice-Boltzmann methods to wall-bounded flows [23]. Ref [24] provides with numerical analysis of bifurcated flow for higher Re numbers for Case B-1 using $100^3$ and $200^3$ grids and second-order of accuracy finite volume code, OpenFOAM.

In Case B-1, the flow does impinge in the spatial angle formed by the front and a side wall. This leads to formation of the system of vortices caused by flow separation. As opposed to the widespread Case C, the flow is substantially three-dimensional for the entire cavity. It is shown in the current study, that helicity in Case A is larger than that in Case B because the velocity in the z direction introduced by moving bottom is parallel to vorticity in the near-lid shear layer moving in the x direction and in the vortex created by moving cavity lid. Consequently, the integral of absolute value of helicity in Case A appears (see Section 3) more than doubled compared to that in Case B-1 for Re=1000.

Per Ref. [25], three-dimensional cavity flow with more than one moving wall does not appear in the literature. Ref [25] extends the 2-D cavity flow in three dimensions so as the top wall is moving to right, while the left vertical wall is moving down with the same constant velocity (see Fig.1, case B-2). In Case B-2, the flows recirculate in upper and lower cavity prisms separated by the cavity diagonal plane that forms plane of symmetry. Each recirculating flow does impinge in the spatial angle formed by the top and left walls. Integral of helicity appears to be comparable to case B-1. In Case B-2, the velocity component in the z direction is close to zero and substantial level of helicity appears only near the edge in which the flows impinge.

Regarding prospective physical experiments to confirm numerical results, lid-driven cavity experiments are described [26], in which moving lid comprises a circularly closed plastic belt driven by a DC motor. Similar apparatus can be established to move bottom wall. To extend to driven cavities with non-planar walls, in experiments [27] investigating lid-driven flow in toroid cavity, the lid rests on the toroid and creates circulating motion in the cavity (see [27], Fig 3.1a). Similar prior set-up [28] is shown in Fig. 3.1b [27]. If the cavity bottom (see Fig. 3.2, Ref [27]) participates in rotational motion, the flow similar to Case A will be obtained.

The study is composed as follows. In Section 2, the mathematical model, boundary conditions, numerical method used, grid convergence and validation of approach are described. In Section 3, flow pathlines and integrals of momentum and helicity for Cases A, B, and C are presented and discussed. Quantitative evaluation of mixing of two fluids in the proposed cavity flow is presented in Section 3 to show that mixing occurs faster than that in the benchmark Case C. In Section 4, local features of the flowfield including primary and secondary vortices are discussed comparatively for Cases A, B and C. Helicity, magnitude of normalized helicity and kinematic vorticity number are evaluated for Reynolds numbers ranging from 100 to 1000. In addition to steady-state flow in driven cavity, the dynamics of vortices is evaluated in response to abrupt change of Reynolds number from 100 to 1000. In Section 5, Conclusions are drawn.

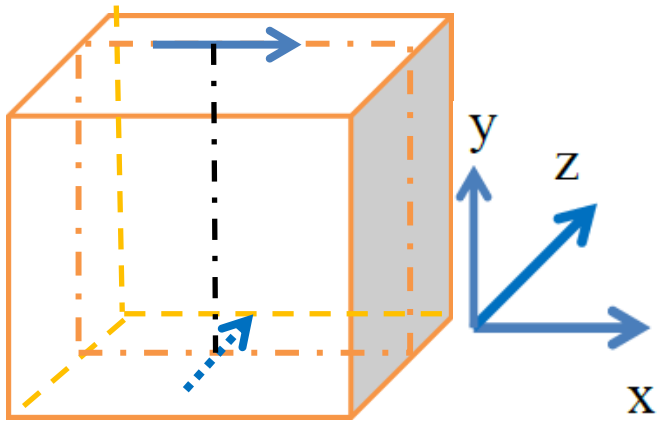
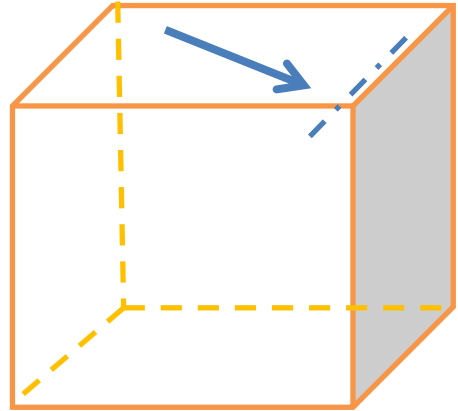

(a)   (b)

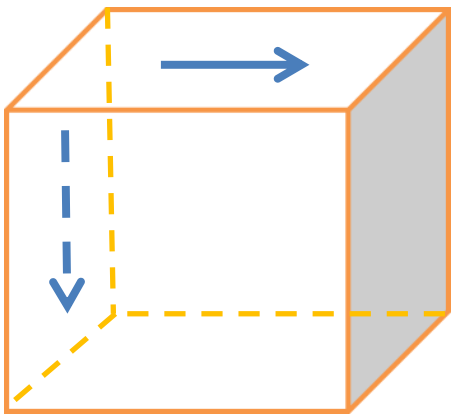
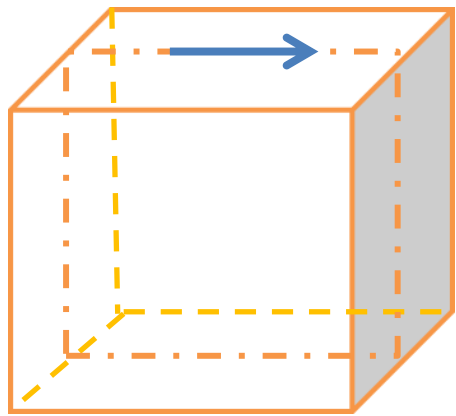

(c)   (d)

Figure 1: Flow in lid-driven cubical cavity: (a) top and bottom walls move in perpendicular directions (Case A), (b) top wall moves along its diagonal (Case B-1), (c) top wall moves along the x direction and the left wall moves down in the y direction (Case B-2), and (d) top wall

moves along its edge (Case C). Central vertical plane (z=0.5), the cavity centerline (x=z=0.5), and near edge line (x=y=0.95) are shown to facilitate presented plots.

2. Numerical method and grid convergence

The governing equations are the three-dimensional steady continuity and Navier-Stokes equations in Cartesian coordinates *(x,y,z)*:

$$\frac{\partial u}{\partial x} + \frac{\partial v}{\partial y} + \frac{\partial w}{\partial z} = 0,$$

(1)

$$\frac{\partial u}{\partial t} + u\frac{\partial u}{\partial x} + v\frac{\partial u}{\partial y} + w\frac{\partial u}{\partial z} + \frac{\partial p}{\partial x} - \frac{1}{Re}\nabla^2 u = 0,$$

$$u\frac{\partial v}{\partial x} + v\frac{\partial v}{\partial y} + w\frac{\partial v}{\partial z} + \frac{\partial p}{\partial y} - \frac{1}{Re}\nabla^2 v = 0,$$

$$u\frac{\partial w}{\partial x} + v\frac{\partial w}{\partial y} + w\frac{\partial w}{\partial z} + \frac{\partial p}{\partial z} - \frac{1}{Re}\nabla^2 w = 0,$$

(2)

where $u,v,w$ are components of velocity in the x, y and z directions, and Laplace operator $\nabla^2 F = \frac{\partial^2 F}{\partial x^2} + \frac{\partial^2 F}{\partial y^2} + \frac{\partial^2 F}{\partial z^2}$.

In the proposed Case A the axis of vortices are in the x and in the z directions that makes the flowfield three-dimensional (see Fig. 4a). The boundary conditions describe the no-slip and non-penetrating conditions at the steady and moving walls. Solutions of Eqs. (1-2) depend on a single parameter named Reynolds number, $Re=\rho UL/\mu$, where ρ and μ are density and viscosity of fluid, respectively, *U* is the speed of cavity lid, *L* is the length of cavity edge. For normalized variables used in the current study *U, L* and *ρ* are taken equal to unity. Consequently, *Re=1 /μ*. Solutions are obtained for *Re=100, 500* and *1000*.

The 3-D cavity flow is studied by numerical solution of the three-dimensional viscous Navier-Stocks equations, Eqs. (1-2), using ANSYS/Fluent finite-volume software with second-order upwind schemes for convective terms ([29] and [31], Section 20.3.1.3) and second-order central scheme for viscous terms. ANSYS/Fluent software uses The Semi-implicit Method for Pressure-linked Equations (SIMPLE) [30] to resolve velocity and pressure coupling in non-linear Navier-Stokes equations. Algebraic multigrid (AMG) ([31], Section 20.7.3) is used to accelerate solution of the Poisson equation for pressure correction. After finite-volume discretization of continuity and momentum equations, conservation equation for a variable ϕ at a finite volume (cell) P can be written [30,31]

$$a_p \phi_p = \Sigma_{nb} a_{nb} \phi_{nb} + b, \quad (3)$$

where $a_p$ is the current finite-volume coefficient, $a_{nb}$ is the coefficient corresponding to a neighboring finite volume and b is a source term.

After finite-volume discretization (3), large system of equations *AX=B* is developed. To reduce the computational time and ensure convergence of iterations, successive under- relaxation iterative method [31] is used to solve the linear system, with relaxation coefficients 0.5 for pressure correction and 0.7 for momentum equations.

The solution for each considered Reynolds number is obtained starting from the zero flowfield, *(u,v,w)=0*. The stopping criteria are based on evaluation of residuals and their comparison to corresponding threshold values. Residual, $R^{\phi}$, is a measure of an iterative solution convergence which evaluates the local imbalance of a conserved variable in each control volume (cell) [31] of the finite-volume grid. In this study, the convergence criteria for discretized mass and momentum equations have been assumed as $R^\phi < 10^{-04}$.

Gauss-Seidel method is used to solve Eqs. (3) for steady-state cases. For transient case, such as transition from steady flow with Re=100 to Re=1000, the implicit second-order temporal discretization method is used ([31], Section 20.3.2) with time step $\Delta t$=0.01.

The uniform numerical grids with grid steps *h=0.005* (*201 x 201 x 201* grid) and *h=0.01* (*101 x 101 x 101* grid) are used. In Fig. 2, profiles of *x*-velocity component u in Case A along the vertical center-line of the cube (*x=z=0.5, 0≤y≤1*, see Fig 1a) are compared for computations with grid steps *h=0.005* and *h=0.01*. The results are quite similar for these two grids. At Re=1000, there is slightly more pronounced local minimum of velocity for more refined grid with *h=0.005*. This is explained by a smaller amount of numerical viscosity added by computations on a more refined grid.

For the benchmark Case C, solution in Fig. 3 practically coincides with prior solutions (see Ref [32], Table 2, Ref. [33], Fig. 7, and Ref. [34], Fig. 10). Solutions [32] were obtained using 50 x 52 x 50 trilinear finite elements. Solutions [33] correspond to prior numerical [35] and experimental [36] results. Ref [33] used uniform mesh up to 101 x 101 x 41 points with symmetric boundary condition. This prior study (see Table 1, Ref [33]) studied grid convergence on coarser grids to accept the 101 x 101 x 41 grid with symmetry boundary condition at z=0.5, which corresponds to 101 x 101 x 81 finite-volume cells for the entire cavity.

Note that the 2-D lid-driven cavity flow (that is, cavity with infinite span in the *z* direction) has qualitatively similar velocity profile; however, the minimum of u component of velocity along the vertical centerline is more pronounced in the 2-D case reaching -0.38 [37, Table 1].

The Dell workstation used in the current study is equipped with Intel Xeon X5660 processor, which has 6 cores/12 threads, and runs at 2.8 GHz with 24 Gigabytes of RAM. The CPU time elapsed using the $200^3$ grid is ~10 seconds per iteration when 10 threads are used. An average run of 2000 iterations till convergence requires ~5.5 hours.

Comparison of u-velocity profile for Cases A and C along the cavity centerline show that profiles are qualitatively similar. In Case A, the maximum of recirculating velocity is located more close to the cavity bottom (that is, y=0) compared to Case C. The upper boundary of recirculating zone (u=0) is more close to the cavity top in Case A. Nevertheless, the z- velocity w introduced by the moving bottom in Case A make the 3-D flowfield quite different compared to that in prior Cases B and C as shown in next sections.

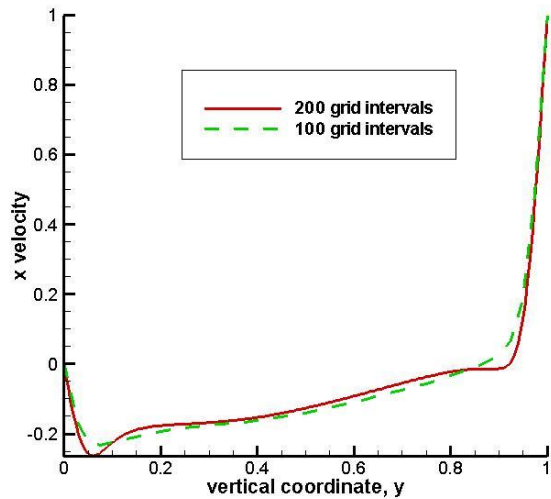

(a)

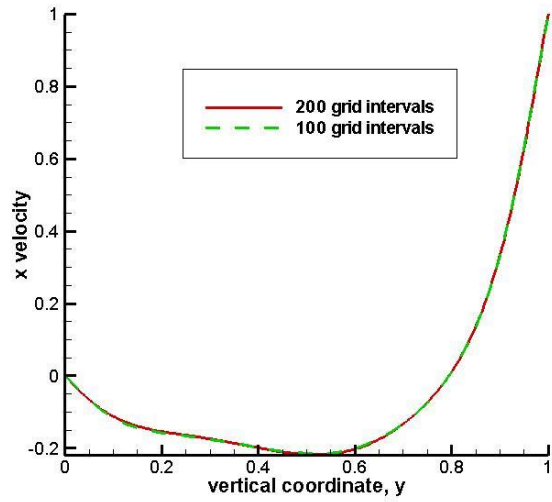

(b)

Figure 2: Distribution of u-component of velocity along the cavity centerline: (a) Re=1000 and (b) Re=100.

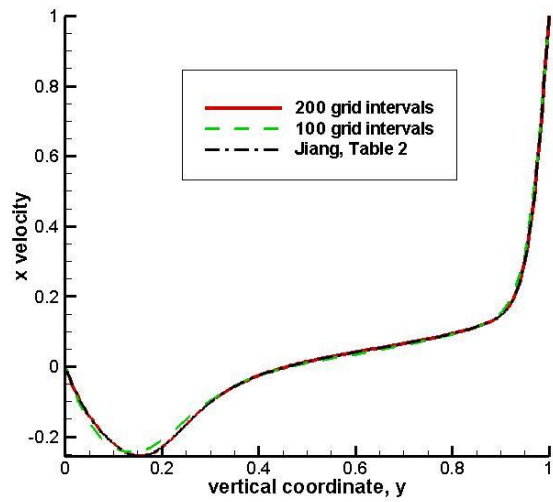

(a)

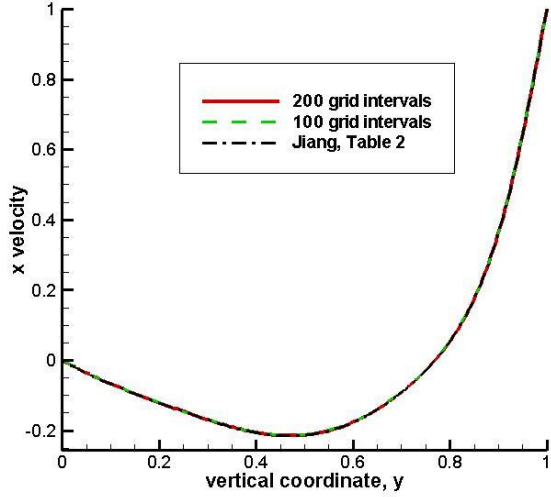

(b)

Figure 3: Benchmark Case C: x-component of velocity along the center line: a) Re=1000 and b) Re=100. Results are compared with Ref. [32], Table 2.

3. Integral properties of flowfield

Table 1 shows the integral of momentum in the x direction for Cases A and C

$$I_T = \frac{1}{V}\int_V M_T, \quad (4)$$

where $M_T = 0.5u^2$. The amount of momentum for Cases A and C is similar. In both cases the integral of momentum somewhat decreases with the increase of Re. The magnitude of helicity (see Table 2) is defined as the integral of the magnitude of helicity density, h, which is the inner product of velocity and vorticity:

$$I_T = \frac{1}{V}\int_V |h|, \quad (5)$$

where $h = \vec{u}\cdot\vec{\omega}$ and $\vec{\omega} = \text{curl}\,\vec{u}$ is vorticity.

Recall that dimensionless quantities are presented in Tables 1 and 2 because length and velocity were normalized.

The magnitude of helicity density is integrated to sum fluctuations of different sign of the local helicity density. The minimum level of helicity is observed in Case C. The maximum level of helicity is in the new proposed Case A. For Re=1000, Cases B-1 and B-2 have more than doubled level of helicity compared to the baseline Case C. It will be shown in the next section that the helicity in Cases B1 and B2 is substantial only at the areas of flow impingement. For Cases B and C there is no substantial velocity in the direction of vorticity.

On the contrary, for the proposed Case A the moving bottom wall introduces the velocity component z that forms the non-zero helicity (Eq. (5)) by the inner product of velocity with vorticity created by the top wall moving in the x direction (Fig 4a). In turn, the moving bottom wall, through the shear stress, generates vorticity oriented in the *x* direction. Multiplying by the velocity component, u, this vorticity component contributes to helicity. As a result, the magnitude of helicity in Case A is ~ 5 times larger than that in Case C if Re=1000 (see Table 2). The difference in magnitude of helicity between Cases A and C does increase with the value of Reynolds number (see Table 2). Details regarding local value of helicity per unit volume and normalized helicity (angle between $\vec{u}$ and $\vec{\omega}$) are presented in the next Section.

Table 1. Momentum in the x direction calculated by Eq. (4)

| Reynolds number | Case A | Case C |
|---|---|---|
| Re=100 | 0.0337 | 0.0374 |
| Re=500 | 0.0260 | 0.0263 |
| Re=1000 | 0.0215 | 0.0253 |

Table 2. Magnitude of helicity calculated by Eq. (5)

| Reynolds number | Case A | Case B-1 | Case B-2 | Case C |
|---|---|---|---|---|
| Re=100 | 0.0778 | 0.224 | 0.0798 | 0.0475 |
| Re=500 | 0.3076 | 0.2025 | 0.1468 | 0.0580 |
| Re=1000 | 0.3899 | 0.1756 | 0.1614 | 0.0798 |

To visualize the flowfield, the pathlines for Case A are depicted in Fig. 4a. To obtain the pathlines using computed velocity vector field, the spatial step size is 0.01 and number of steps is 500. The total 1600 particles are traced; of them the first 800 particles are ejected from the bottom surface and the remaining 800 particles are ejected from the top surface. These pathlines are colored by the particle number. Two primary vortices are clearly seen in Fig.4a: the upper one with its axis oriented in the z direction and the lower one with its axis oriented in the x direction. Pathlines created by particles ejected from top and bottom surfaces are mixed well with each other: red and yellow pathlines corresponding to particles ejected from the top surface are mixed with blue and green pathlines corresponding to particles ejected from the bottom surface. Particles, which are ejected from bottom and top walls, participate in the swirling motion to the same degree. This indicates the uniform involvement of material into swirling motion and mixing in Case A.

The pathlines in Case B-1 are depicted in Fig. 4b. The vortex tube is oriented in the (x,-z) direction parallel to the direction of moving lid (see Fig. 1b). The vortex tube is bended so as it nearly touches the upper wall at its edges and is deflected down toward the cavity center. The bended shape of vortex tube creates numerous secondary vortices of lesser strength as detailed in the prior study by the author [18]. The first 1600 particles are ejected primarily from the bottom and the other 1600 particles are ejected from the top. The blue and green color of pathlines indicates that most of material involved in swirling motion is ejected from the bottom.

In Case C, pathlines form primary vortex with its axis oriented in the x direction. Yellow and red colors of pathlines indicate that most of material involved in swirling motion is ejected from the top with the exception of blue pathlines of particles ejected from bottom, which enter the vortex through its edges and move toward center of cavity.

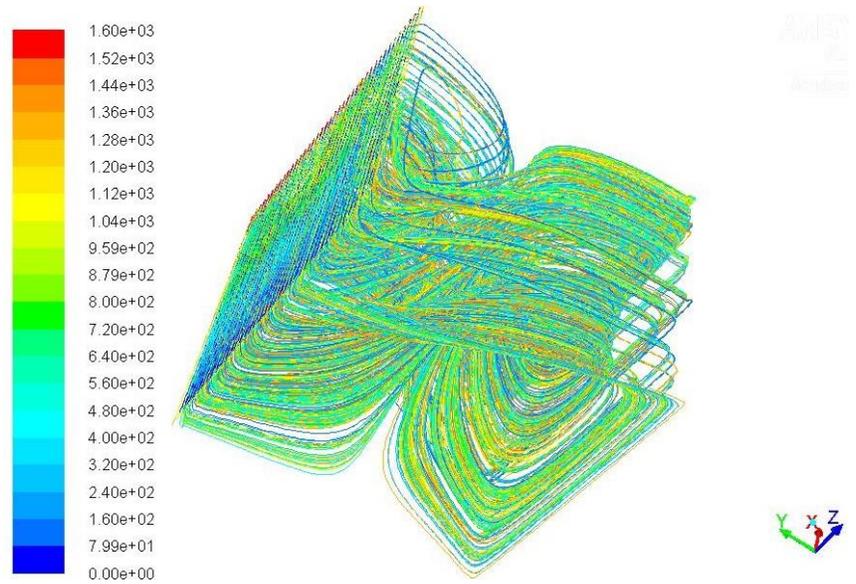

(a)

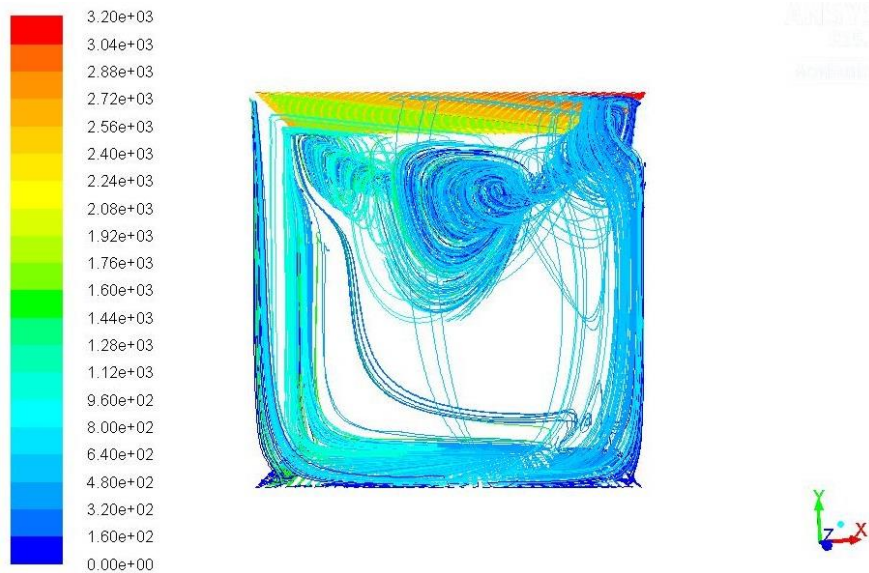

(b)

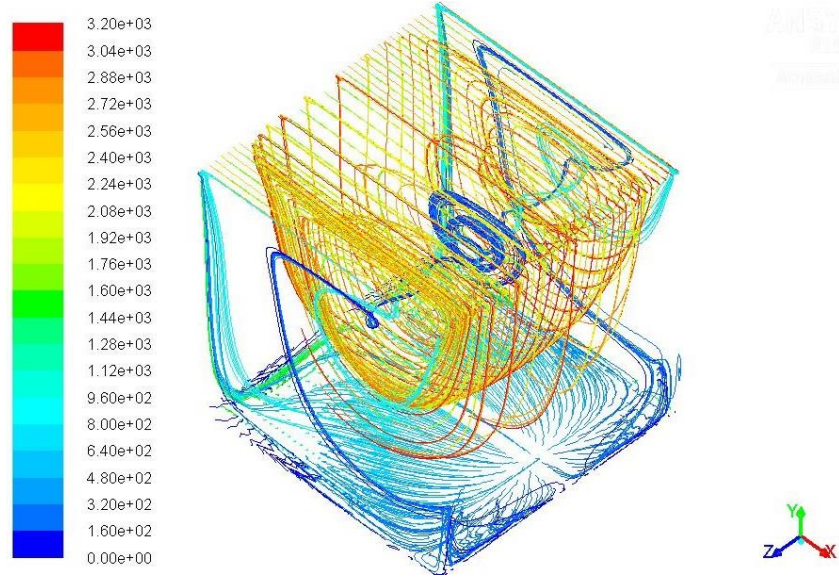

(c)

Figure 4: Pathlines colored by particle number ejected from bottom and top surfaces: (a) Case A, (b) Case B-1, and (c) Case C.

To evaluate mixing quantitatively, the steady-state cavity flowfield established above is filled with two fluids of the same density and viscosity. The first fluid occupies the upper half of cavity (y>0.5) while the second fluid fills the lower half of cavity (y<0.5). Therefore, the plane y=0.5 divides two species with the same properties and the average value of concentration of each fluid in cavity is 0.5. The dispersion (degree of non-mixing of these fluids) can be quantified by value of the mean square variable

$$\sigma^2 = \int_V (c - 0.5)^2, \quad (6)$$

where c is the mass concentration of the first fluid. The value of $\sigma^2$ is equal to $\sigma_0^2 = 0.25$ at t=0. For t→∞, the variable $\sigma^2$ tends to zero, that is, both fluids are perfectly mixed after long-time mixing. The value of normalized mean square variable, $\sigma^2 / \sigma_0^2$, is shown in Fig.

5 as a function of unit-less time $T=tU/L$. For the benchmark Case C, the value of $\sigma^2 = 0.0351$ at $T = 10$. For the proposed Case A $\sigma^2 = 0.01324$ at $T = 10$, that is, 2.65 times smaller than that in Case C. This indicates faster mixing in the proposed Case A. For Cases A and C, contour plots of variable c in central vertical plane (see Fig. 1) are presented in Fig. 6a, b, respectively. Fig. 6 shows that the maximum value of concentration is bigger in Case C compared to Case A. Area of maximum concentration (that is, of non-mixed species) is bigger in size for Case C and occupies entire upper left corner. For Case A, the maximum concentration occupies smaller area in upper left corner. For Case A, the small patches of high concentration of the first fluid appear near bottom, where it was zero initially, that confirm two-sided mixing.

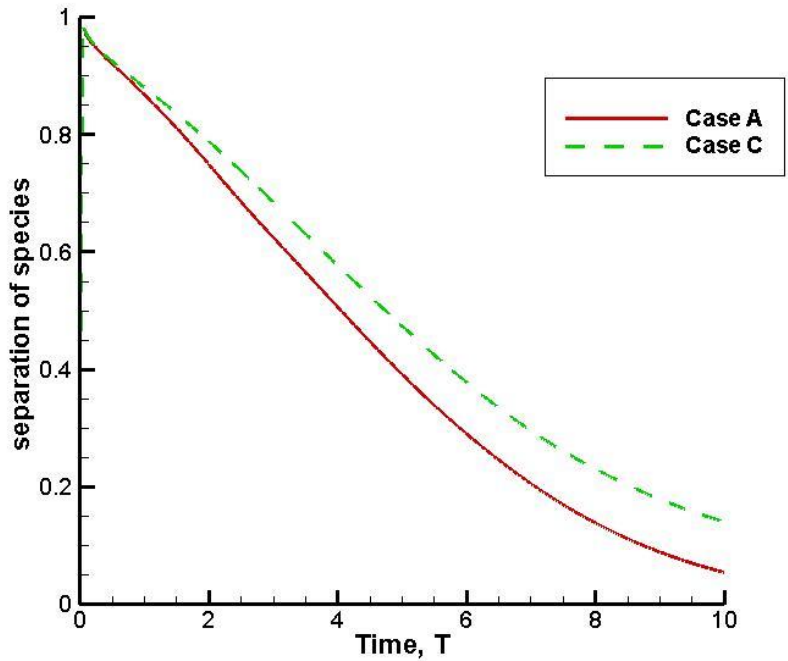

Figure 5: The degree of separation of species ($\sigma^2/\sigma_0^2$) as a function of unitless time, T, for Re=1000.

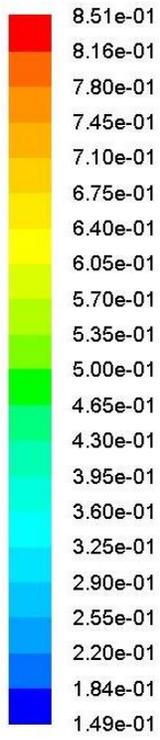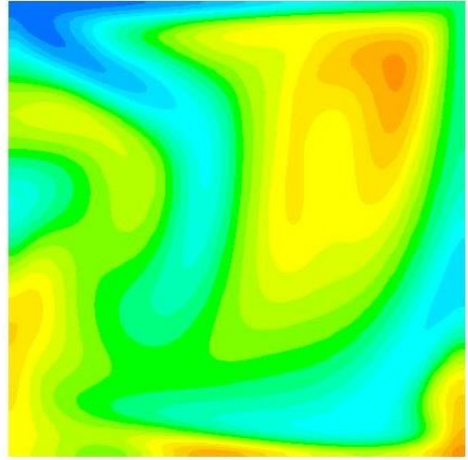

(a)

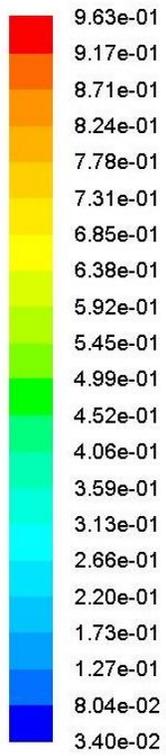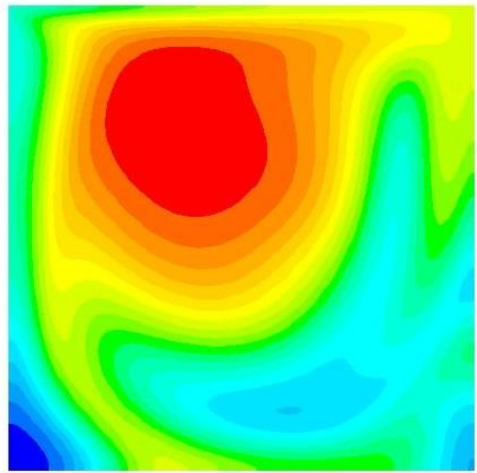

(b)

Figure 6: Contour plots of concentration in central vertical plane at Re=1000 for (a) Case A and (b) Case C. Plots are presented at time moment T=10 after adding two separated fluids in the steady flowfield.

4. Local flowfield and distribution of helicity

In Figure 7a, b, z-components of velocity and vorticity are plot along the cube centerline (see Fig. 1a). The centerline is located outside of vortex tubes (see vertical cross-sectional vector field, Fig. 8); therefore, the distribution of helicity is dominated by shear layers caused by moving top and bottom walls. The z component of velocity, w, along the cube centerline is a mirror reflection of the u component with respect to y=0.5 (compare Fig. 7a and Fig. 2a). The recirculating zone (w<0) is formed in the upper part of the cavity. For Re=100, the near-wall layers dominated by moving walls are somewhat wider compared to those for Re=500 and 1000.

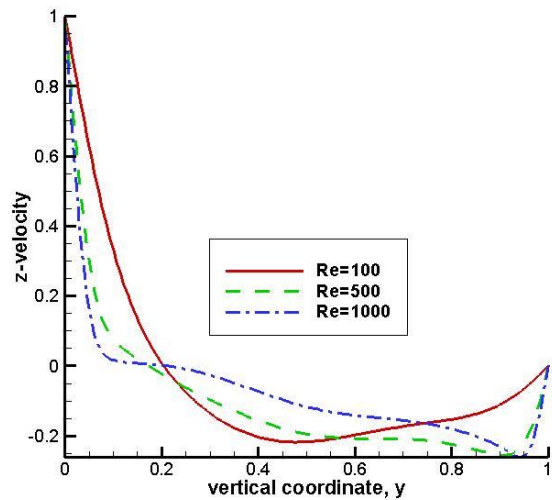

(a)

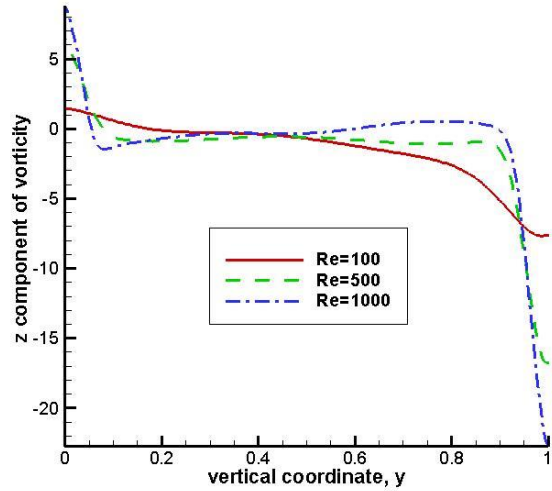

(b)

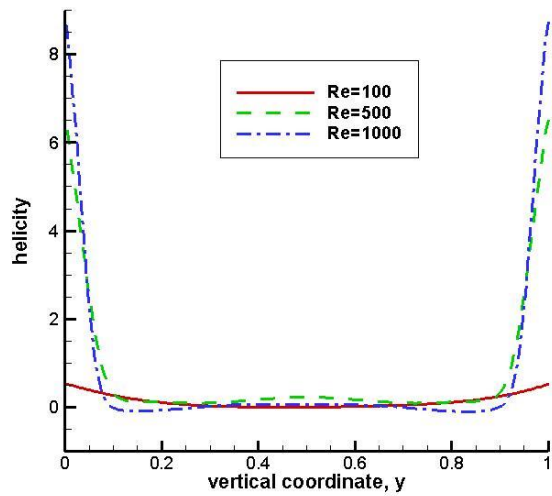

(c)

Figure 7: Case A: centerline (*x = z = 0.5*) profiles of (a) z component of velocity, (b) z-component of vorticity and (c) magnitude of helicity

Velocity vector field colored by helicity density, h, is plot in vertical cross-section (z=0.5) and shown in Figs. 8a, b, c for Re=100, 500 and 1000, in correspondence. The velocity vector flowfield for Case C (Re=1000) is presented in Fig. 8d for comparison. The primary vortex is created by the top of the cavity moving in the positive x direction. The shear stress is applied to the upmost layer of fluid that causes its motion in the x direction. The right wall blocks the forward motion of fluid and turns it toward the bottom in Case C. In Case A, formation of vortex by motion of the bottom wall in the z direction deflects the bottom-moving fluid toward the diagonal of cross-section.

In Case A (Re=500 and 1000), the secondary co-rotating vortex is formed above the impinging flow near the lower part of right wall, see Fig. 8 b, c. For Re=100 (Fig. 8a), the low velocity magnitude flow region is formed in place of secondary vortex.

In Case C, the fluid moving toward bottom form a secondary counter-rotating vortex near the bottom right corner (compare Figs 6a,b,c to Fig. 8d). In Case C (Fig. 8d), the geometry is an extension to three dimensions of the 2-D lid-driven square cavity in which the secondary vortices are located at near-bottom corners. When the Reynolds number increases, the corner eddies grow larger [38].This vortex is not observed in Case A (Fig 6a, b, c) being convected downstream by the flow in the z direction.

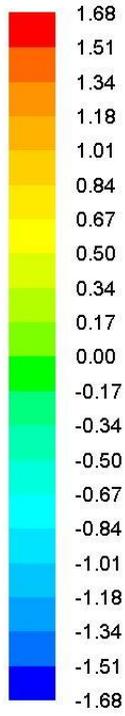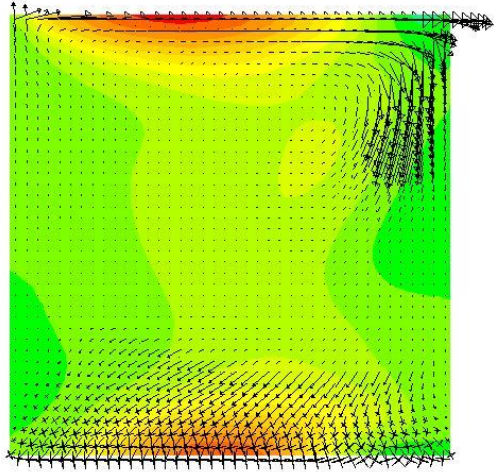

(a)

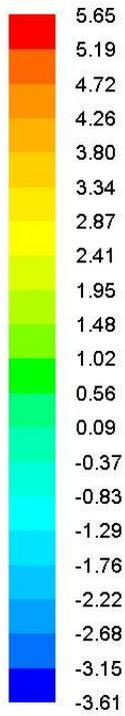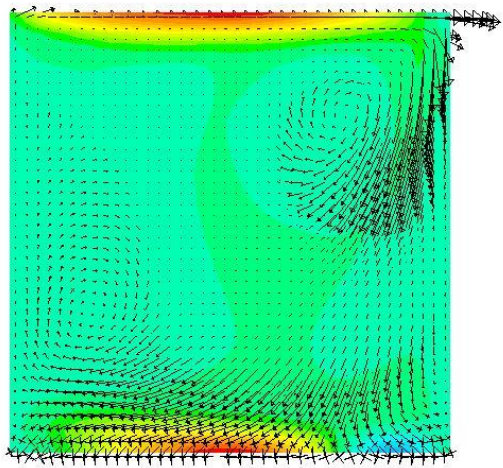

(b)

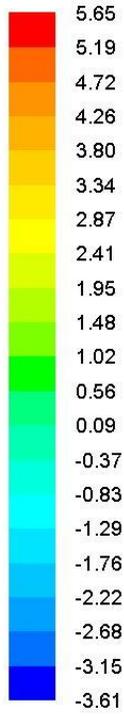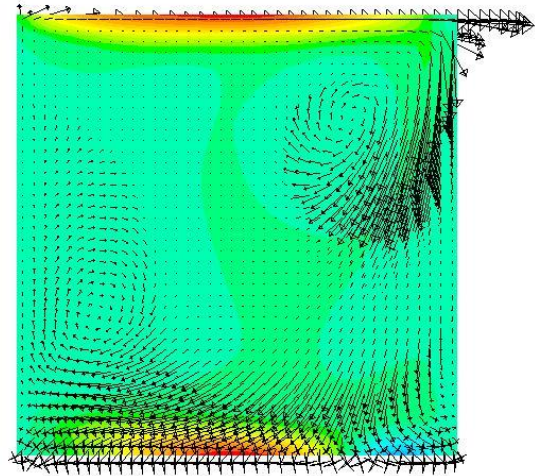

(c)

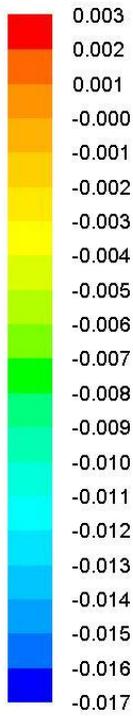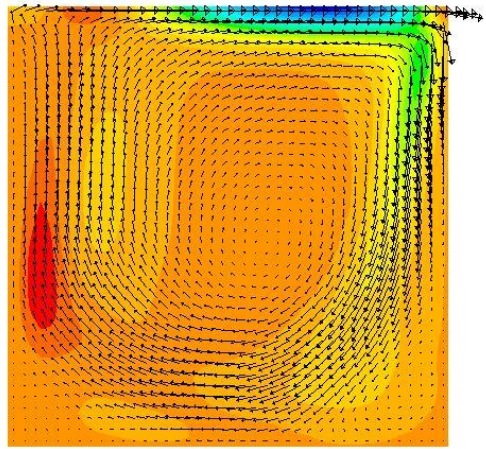

(d)

Figure 8: Velocity vector field and contours of helicity density, h, in vertical cross-section (z=0.5) for Case A (a) Re=100 (h from -1.68 to 1.68) (b) Re=500 (h from -3.61 to 5.65), (c) Re=1000 (h from -3.61 to 5.65), and for Case C (d) Re=1000 (h from -0.017 to 0.003)

As soon as velocity vector field and topology of vortices are different for Re=1000 compared to Re=100 (Fig. 8a, c), the results of transient computations are presented in Fig. 9. The solution was steady-state for Re=100 at T=0. At this time moment the value of viscosity changes so as Re increases from 100 to 1000.

At T=1 (Fig. 9a), the velocity magnitude increases in flow impinging in the lower right corner where the secondary vortex is yet to be formed. At T=5 (Fig. 9b), the secondary co-rotating vortex is formed near the left wall close to y=0.5. The primary vortex moves toward upper right corner. At T=10 (Fig. 9c), the secondary vortex moved down and the topology of vortices become similar to its steady state counterpart at Re=1000 (see Fig. 8c).

In addition to velocity field, vorticity and helicity, other quantitative criteria for identification of vortices are used. Review studies of vortex identification for 2-D [39] and 3-D [40] cases compare identification methods: Δ, Q, $\lambda_{ci}$, and $\lambda_2$ criteria are derived and compared mathematically and experimentally. In particular, the Q-criterion by Hunt [41] evaluates the ratio of magnitude of vorticity, Ω, and rate-of-strain, S, $Q = 0.5(|\Omega|^2 - |S|^2)$. By [39,40], listed above vortex identification criteria are quite sensitive to the chosen threshold values for the vortex analysis.

The kinematic vorticity number, $W_k = \frac{|\Omega|}{|S|}$, was introduced by Truesdell [42] to indicate the amount of rotation relative to the amount of stretching, at a point in space and in an instant in time (see [43, 44, 45] and references therein). The Q –criterion is related to vorticity number, $W_k$ [46]. As opposed to the Q-criterion, $W_k$ has clearly defined threshold values [45]. If W{k} > 1, rotation rate prevails over strain rate.

The $W_k$ criterion appears to be more suitable for wall-bounded flows compared to Q-criterion. For wall-bounded flows the rate-of strain, S, dominate near the walls and vortices are not as clearly seen by Q-criterion as by the W{k} criterion, that is plot in Fig. 9. The area corresponding to near-maximum values of the W{k} also corresponds to location of vortices by velocity vector field.

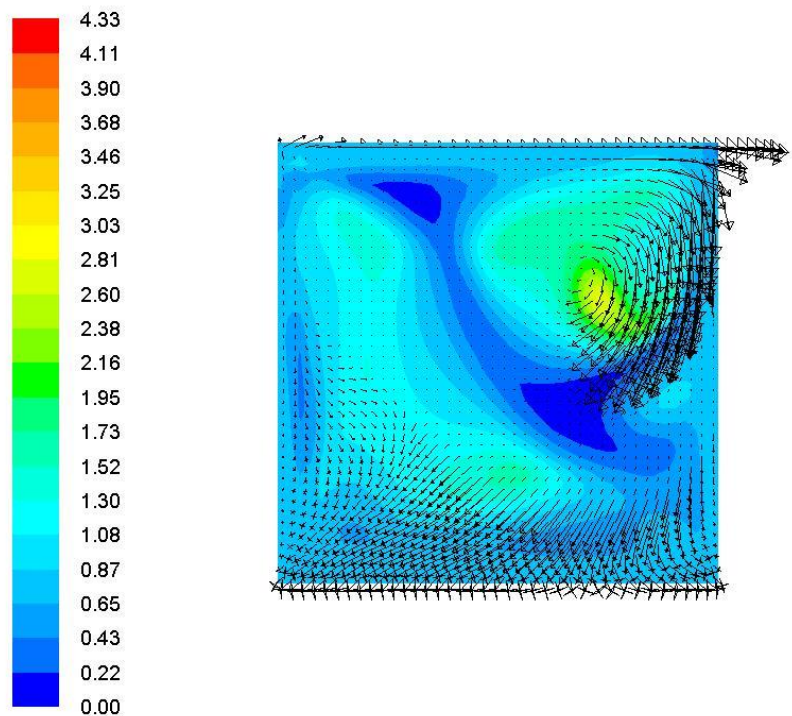

(a)

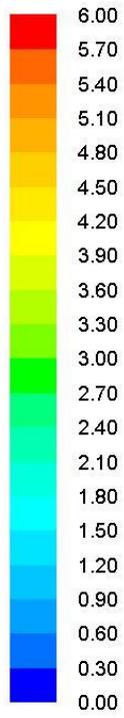 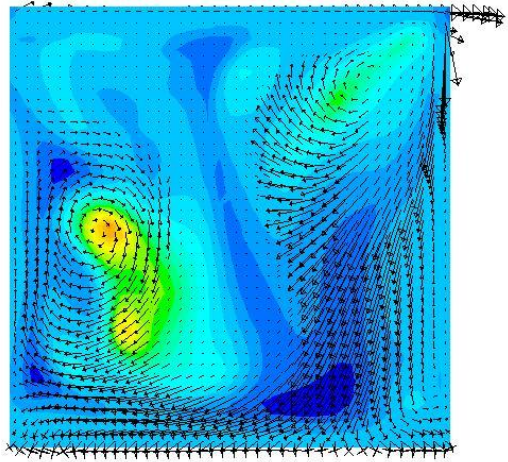

(b)

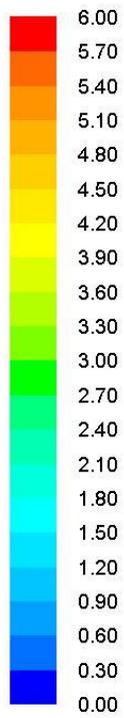 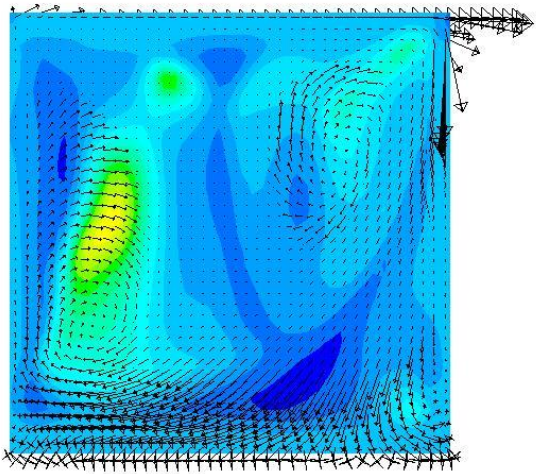

(c)

Figure 9: The contour plots of kinematic vorticity number, $W_k$, and velocity vector field in the central vertical plane: (a) T=1, (b) T=5, and (c) T=10. At time moment T=0 the value of viscosity changes so as Re increases from 100 to 1000.

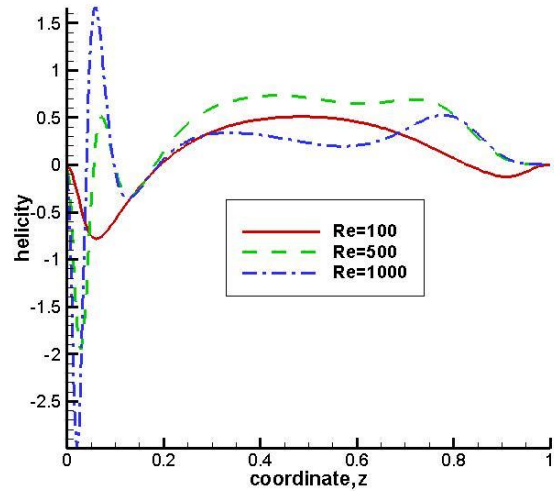

(a)

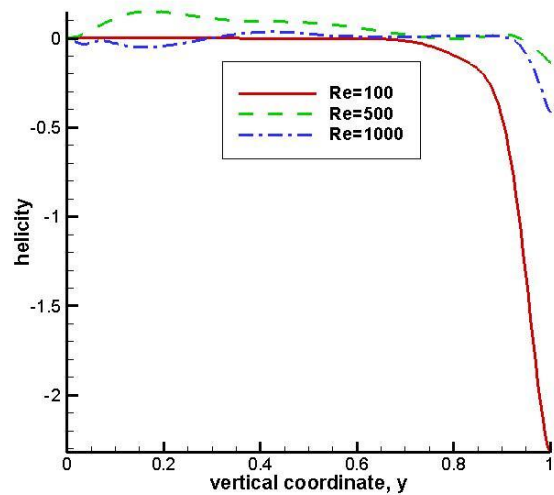

(b)

Figure 10: Helicity density in Case B-1: (a) along near-edge line (x=y=0.95) and (b) along the vertical centerline (x=z=0.5)

In Fig. 10a, helicity density, h, is plot along the horizontal line (x=y=0.95) parallel to upper right cavity edge (Fig. 1b). In Case B-1, the lid-driven flow impinges into upper right corner (z=0) in Fig. 1b. The magnitude of helicity in Fig. 10a is below 0.5 except for impingement area (z<0.1) at which the absolute value of h reach ~3 for Re=1000. In Fig. 10b, helicity is plot along the cube centerline in Case B-1. Maximum magnitude of helicity near the cavity top is ~0.3 for Re=1000.

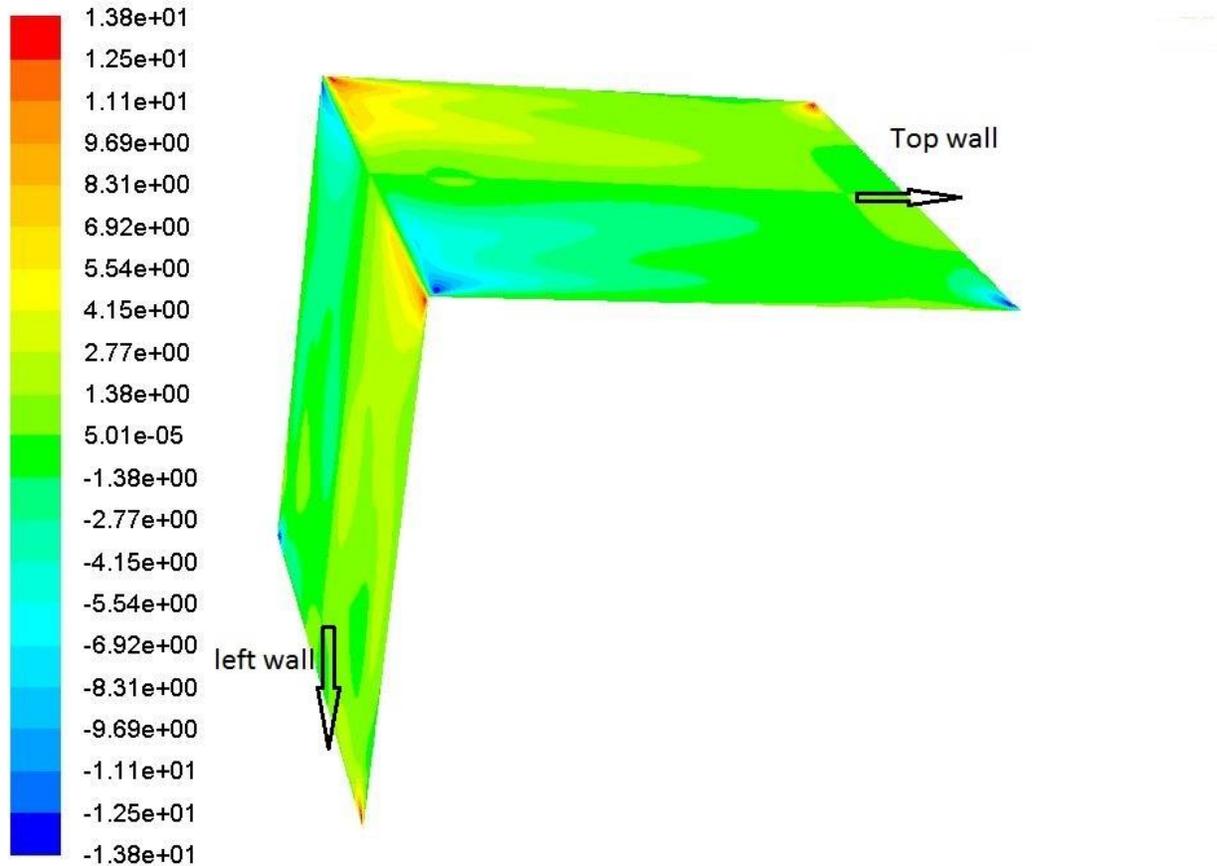

Figure 11: Helicity in Case B-2

In Figure 11, helicity density, h, is plot at the top and left cavity walls to show the areas of elevated helicity in Case B-2. For the rest of the cavity flow the helicity is close to zero except for areas located near the corners about which left and top walls move. In summary, the level of helicity in Cases B-1 and B-2 are much smaller compared to Case A (see Fig. 7c).

The degree of alignment of the velocity and vorticity can be evaluated using the normalized (relative) helicity [11]:

$$\cos(\alpha) = h/(|\vec{u}| \cdot |\vec{\omega}|) \qquad (7)$$

The value of $\cos(\alpha)$ is close to 1 when vectors of velocity and vorticity are parallel and is close to -1 when they are antiparallel. The value of $\cos(\alpha)$ is close to zero when velocity and vorticity are non-aligned. Flow situations in which the magnitude of normalized helicity is close to unity is important in dynamics of vortices and turbulence [9]. The flowfield in Case A naturally has such areas of flowfield because the motion of each one of walls (top or bottom) creates velocity field aligned with the vortex axis created by another parallel wall.

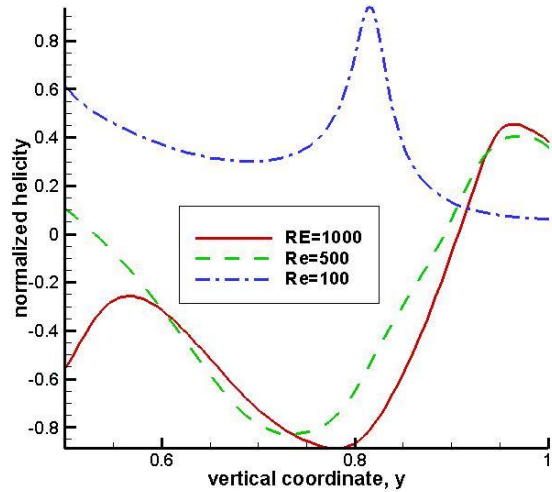

Figure 12: Normalized helicity in Case A. Normalized helicity is plot along vertical lines (z=0.5) that passes through the vortex center (x=0.65 for Re=1000, x=0.72 for Re=500, and x=0.626 for Re=100)

In Figure 12, the normalized helicity, $\cos(\alpha)$, is plot along vertical lines, 0.5≤y≤1. Each line passes through the vortex center (x=0.65 for Re=1000, x=0.72 for Re=500, and x=0.626 for Re=100) at the middle of the cavity (z=0.5). The graphs in Fig. 12 show that the magnitude of normalized helicity $|\cos(\alpha)|$, exceeds 0.8 at the central part of each vortex. The sign of normalized helicity is opposite for Re=100 and for larger Reynolds numbers (Re=500 and Re=1000) because the sign of vorticity is opposite for these cases, see Fig. 7b. The normalized vorticity plots show that the maximum magnitude of normalized helicity is reached at core of vortices while the maximum magnitude of vorticity (Fig. 7b) and helicity density (Fig. 7c) correspond to near-wall layer.

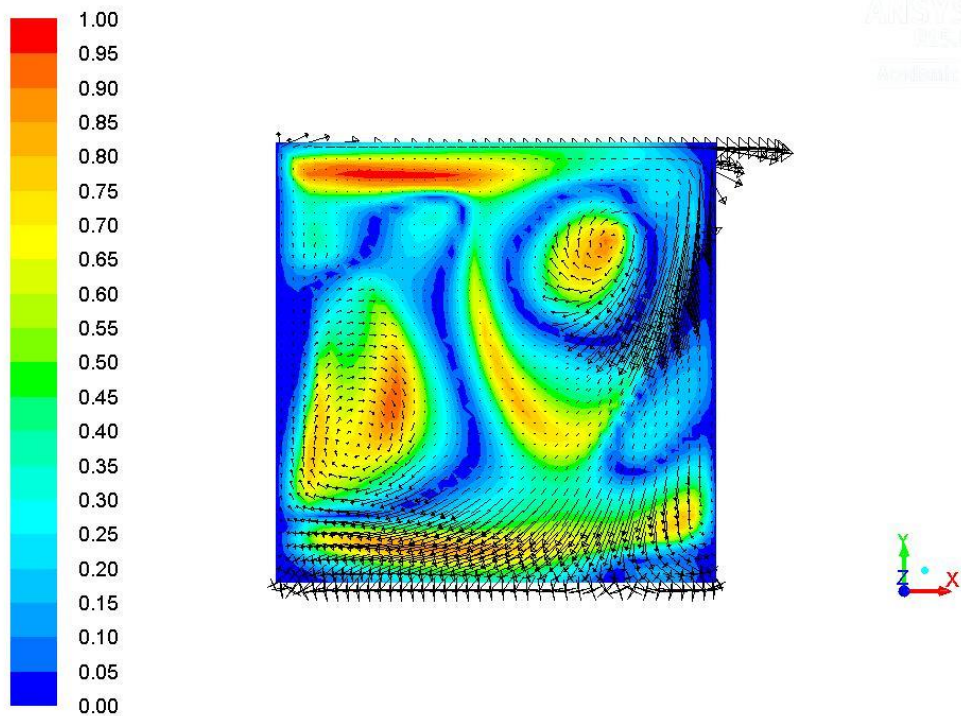

Figure 13: Magnitude of normalized helicity, $|\cos(\alpha)|$, at central vertical plane for Case A, Re=1000

To show elevated level of normailzed helicity at vortices, the velocity vector field and magnitude of normalized helicity are shown at central vertical plane in Fig. 13. The primary and secondary vortices do correspond to an elevated level of magnitude normalized helicity, that is, $\alpha$ is either 0 or 180 degrees.

Conclusions

The proposed flow in 3-D cubic cavity is driven by its top and bottom walls moving in perpendicular directions. This flow is a representative of genuinely three-dimensional highly separated vortical flows yet having simple single-block cubical geometry of computational domain.

To obtain the flowfield, the governing incompressible Navier-Stokes equations are solved numerically using second-order upwind scheme for convective terms and second-order central scheme for viscous terms. Solutions obtained on $100^3$ and $200^3$ grids show close proximity to each other for both proposed Case A and benchmark Case C. The proposed flow is also compared to flows characterized by a diagonal plane of symmetry (Cases B-1 and B-2). To validate the numerical method, solutions in Case C appear to be identical to those known in literature. The $200^3$ grid is therefore used for numerical modeling in the current study.

The streamlines for the proposed flow show formation of two primary vortices with perpendicular axis. The particles emerging from top and bottom appear to mix well with each other. The proposed flow is different from the well-known benchmark Case C in which the particles emerging from the top were primary involved in circulating motion and from the Case B-1 in which the particles emerging from the bottom were chiefly involved in circulating motion. The rate of mixing of initially separated fluids was evaluated quantitatively by tracing the average over cavity deviation of local concentration of first fluid from its average concentration. The rate of mixing for the proposed flow appears to be substantially faster than that in the benchmark Case C.

For proposed flowfield, primary and secondary vortices, which are created for Re>100, are located quite differently within the cavity compared to those in the benchmark Case C. Dynamics of patterns of vortices is evaluated by tracking the transient Case A flow in response to abrupt change of Reynolds number from 100 to 1000. The primary vortex moves from its near-central location in cavity at Re=100 toward the upper right corner. The secondary vortex is formed in the stagnation zone in the lower left quadrant of cavity when the value of Re exceeds 100. The forming secondary vortex moves in vertical direction before settling when Re approaches 1000. The kinematic vorticity number appears to be a good indicator of location of vortices in cavity.

Helicity is quantified as an inner product of vorticity and velocity and is considered as an important measure of transition to turbulence and degree of mixing. The elevated level of helicity in the proposed flowfield is caused by the motion of a wall (either top or bottom) in the direction of vorticity created by another moving parallel wall. The vorticity is created by viscous shear stress caused by moving wall and by vortex created by no-penetration effect of stationary side walls, which block the shear flow in the direction of moving wall. For Re=1000, helicity is five times bigger for the proposed flow compared to flow driven by the cavity lid (Case C). For cavity flows driven by lid moving along its diagonal (Case B-1) and by two perpendicular walls (Case B-2), the helicity is approximately two times smaller compared to Case A. For Case B flows, the elevated helicity level is limited to the corners of the cavity where the driven flow impinges into walls. For the proposed Case A flow, the helicity is elevated in the spatial areas next to moving top and bottom walls (y<0.1 and y>0.9) while the magnitude of normalized helicity (cos of angle between $\vec{u}$ and $\vec{\omega}$) is close to unity within flow vortices.